\documentstyle[12pt,epsfig]{article}                 
\begin{document}
\begin{titlepage}
\begin{flushright}
TIFR/TH/98-31 \\
UTCCP-P-42 
\end{flushright}
\vskip1.5cm

\begin{center}
{\Large {\bf Re-assessing the anomalous $J/\psi$-suppression in the CERN
NA50 data}} \\[15mm]

{\large Rajiv V. Gavai$^{a,b,}$\footnote{E-mail:gavai@tifr.res.in} }  \\[5mm]
{\em $^a$Theoretical Physics Department, Tata Institute of Fundamental
Research, \\
Homi Bhabha Road, Mumbai 400005, India} \\[2mm]
{\em $^b$Center for Computational Physics, University of Tsukuba, \\
Tsukuba, Ibaraki 305-8577, Japan} \\[20mm]
\end{center}

\begin{abstract}

A systematic analysis of the $L$-dependence of the $J/\psi$-suppression
in the data of the CERN NA38 and NA50 experiments shows that the
anomalous suppression in the 1995 $Pb{\rm -}Pb$ data is at best a 
4$\sigma$ effect at any of the $L$-values for the $Pb{\rm -}Pb$ data, where 
$L$ is the geometrical mean path length of the $J/\psi$ in the colliding 
nuclei. Possible implications for the 1996 data are discussed.

\end{abstract} 

\end{titlepage}

Discovery of Quark-Gluon Plasma (QGP) is one of the most exciting aims
of the current heavy ion collision experiments for a variety of
reasons.  Unfortunately,  however, the evaporation of QGP is
expected to occur in such a short time that establishing its 
formation by distinguishing possible hadronic  backgrounds to
the proposed signals of QGP is a very non-trivial task.
Observation of  $J/\psi$-suppression\cite{MatSat} 
has been thought of as a particularly promising way of 
looking for QGP.  Here too, however, nuclear structure function
effects\cite{GGS} and absorption\footnote{ It has been 
argued\cite{KhaSat} that the usual color singlet $J/\psi$ has 
too small absorption cross section and it is the color octet pre-resonant
$c \bar c$ which contributes to this mechanism.  We will not need 
to worry here about such a distinction.} of produced $J/\psi$'s by
the surrounding nucleons\cite{GerHue} or the co-movers\cite{GavGyu}
are additional competing mechanisms.

The recent announcement\cite{Qm97} by the CERN NA50 experiment of 
observing once again anomalous $J/\psi$-suppression in their more precise
1996 data assumes a lot of significance in view of their earlier
published results\cite{Qm96,Na50Plb} where they obtained anomalous
suppression at about  5$\sigma$ level in the total $J/\psi$
cross section in Lead-Lead collisions at 158 GeV/A and at about 10$\sigma$ 
level in the $L$-dependence of the suppression at their largest $L$-values.
Here $L$ is the geometrical mean path length of $J/\psi$ in the colliding
nuclei.  It is obtained from the measured transverse energy $E_T$
through its relation with the impact parameter $b$ by performing a
weighted average over possible production points of $J/\psi$.
Proton-nucleus data and nucleus-nucleus data for light projectiles 
provides the standard in both the cases with respect to which
the anomalous suppression is measured.  As pointed out earlier\cite{GG},
there are various theoretical uncertainties in this way of estimating
the anomalous suppression.  These range from simple propagation
of errors due to the fitting procedure used to parameterize the usual
suppression to the more intricate uncertainties in scaling some
observed cross sections  to energies other than the measured.
Ref. \cite{GG} has evaluated many of these uncertainties for the
anomalous suppression in the total cross sections and concluded that
there was no anomalous suppression at 95 \% confidence level, i.e.
it was a less than a 2$\sigma$ effect.  The experimental effect is 
claimed to be a lot stronger in the $L$-dependence of the suppression.  
In this brief note, we re-assess its statistical significance and find that
the anomalous suppression at any given $L$ is at most a 4$\sigma$
effect.  We also update the results of Ref. \cite{GG} in view of
the changes in the published\cite{Na50Plb} 1995 data of NA50
compared to their preliminary results\cite{Qm96}.

In its data analysis\cite{Na50Plb}, the NA50 collaboration fits the
the (rescaled) $pA$ data and $A^\prime A$ data to a power-law
\begin{equation}
B\sigma({\cal A})\;=\;\sigma_0 {\cal A}^\alpha~~,
\label{fit.form}\end{equation}
where $B$ is the branching ratio for $J/\psi\to\mu^+\mu^-$, and
${\cal A}$ is the effective mass number, given by the product of
the mass numbers of the (light) projectile, $A^\prime$, and target $A$.
The $L$-dependence of the ratio $R_{\rm expt.}$ of the $J/\psi$ 
cross section to the Drell-Yan cross section is, on the other hand, fitted by
\begin{equation}
R_{\rm expt.} \;=\; C \exp ( - \rho_0 \sigma_{abs} L)~~,
\label{fit1.form}\end{equation}
where $C$ is a constant, $\rho_0 = 0.17~{\rm fm}^{-3}$ is the nuclear matter density
and $\sigma_{abs}$ is the absorption cross section for the $J/\psi$
in nuclear matter.  By taking logarithms of both equations, the fits can be 
thought of as straight line fits.  However, the NA50 data has measurement 
errors on {\it both} $L$ as well as the ratio $R_{\rm expt.}$; the former
come from binning of events in transverse energy. As a consequence,
a straightforward least squares fit is inadequate in this case,
unlike equation (\ref{fit.form}).  One, therefore, has\cite{NumRec}
to minimize the $\chi^2$, defined by
\begin{equation}
\chi^2(a,b)\;=\;\sum_{i=1}^N {{(y_i - a - bx_i)^2} \over
{\sigma^2_{yi} + b^2 \sigma^2_{xi} }}~~,
\label{chisq}\end{equation}
where $y(x)=a +bx$ is a straight line fit to a data set of $N$ points
$(x_i, y_i)$, $i$=1, $N$, with $\sigma_{xi}$ and $\sigma_{yi}$ as the
standard deviations in the $x$ and $y$ directions respectively.

The usual statistical procedure\cite{intro} to propagate errors is the
following.  Let us assume that the expectation values of a
set of variables, $p_i$, are known along with their full covariance matrix,
$c_{ij}$.  Since we deal here with an $f(p_i)$ which is a linear function of 
the $p_i$, $\langle f(p_i)\rangle= f(\langle p_i\rangle)$, and the 
error
\begin{equation}
   (\Delta f)^2\;=\;\sum_{ij} c_{ij}{\partial f\over\partial p_i}
                                    {\partial f\over\partial p_j}~~.
\label{stat}\end{equation}
Note that when the correlations vanish, this reduces to the usual
formula for adding errors in quadrature.   Obtaining the covariance
of $a$ and $b$ in a linear fit is simple but it could be numerically
tricky in the case of $\chi^2$-minimization for equation (\ref{chisq}).
One can overcome this by a simple observation\footnote{I thank Ruedi
Burkhalter for pointing this out to me.}.  Making a transformation
$x^\prime = x + x_0$, changes the intercept from $a$ to $a^\prime = 
a + b x_0$.  Thus getting an error estimate on the parameter $a^\prime$
for the shifted data set under this transformation  is the same as 
obtaining the $\Delta y(x_0)$ including the full covariance matrix.
We use this method to obtain $\Delta y$ at each point $x_0$ by
defining it as usual as the variation in $a^\prime$ which produces a change
in $\chi^2_{\rm min}$ by one.

Using equation (\ref{fit.form}) to fit the latest\cite{Na50Plb}
$pA$ data, one obtains 
\begin{equation}
   \sigma_0=2.28(1\pm0.07)\ {\rm nb},\quad
   \alpha=0.91\pm0.02,\quad
   {\rm Cov}(\log\sigma_0,\alpha)=-0.0013~,
\label{fit.vals}\end{equation}
leading to a prediction 
$B\sigma(Pb{\rm -}Pb)=0.88(1\pm0.13)$ nb. The measured 
point\cite{Na50Plb}, $B\sigma(Pb{\rm -}Pb)=0.67\pm0.05$ nb, is,
therefore, within 2$\sigma$ of the extrapolation.  Adding further the dataset 
with light nuclei projectiles by assuming that  no other source of suppression
than the one operative in $pA$ collisions exists in those cases too,
one obtains,
\begin{equation}
   \sigma_0=2.26(1\pm0.06)\ {\rm nb},\quad
   \alpha=0.914\pm0.013,\quad
   {\rm Cov}(\log\sigma_0,\alpha)=-0.0007~,
\label{fita.vals}\end{equation}
leading to a prediction for the $Pb{\rm -}Pb$ cross section 
as $0.904\pm0.078$ nb, which again is within 2$\sigma$
of the measured cross section.   The smaller errors on the $pA$ and 
$A^\prime A$ data\cite {Na50Plb}, thus result in predictions with slightly 
better error estimates compared to the original results of Ref.
\cite{GG}.  However,  the final measured 
$Pb{\rm -}Pb$ cross section has also moved up a little so that the 
measured cross section lies within 2$\sigma$ of the prediction irrespective
of the inclusion or exclusion of the light nuclei $A^\prime A$ data
in the fit.

As mentioned earlier, the $L$-dependence of the $J/\psi$ cross section
yielded\cite{Na50Plb} a statistically lot more significant result
than the 5$\sigma$ for the case discussed above.  Using 
equation (\ref{fit1.form}) in this case along with the $\chi^2$
as defined in equation(\ref{chisq}), we find
\begin{equation}
   C =45.2(1\pm0.13 ),\quad
   \sigma_{abs} =6.64\pm1.11~{\rm mb}~~.
\label{fitl.vals}\end{equation}
If we set, by hand, all errors on $L$ to zero, then we get 
\begin{equation}
   C =43.3(1\pm0.07 ),\quad
   \sigma_{abs} =6.28\pm0.66~{\rm mb}~~,   
\label{fits.vals}\end{equation}
which is in excellent agreement with the results in \cite{Na50Plb}.

\begin{figure}[htbp]\begin{center}
\epsfig{height=12cm,width=12cm,angle=-90,file=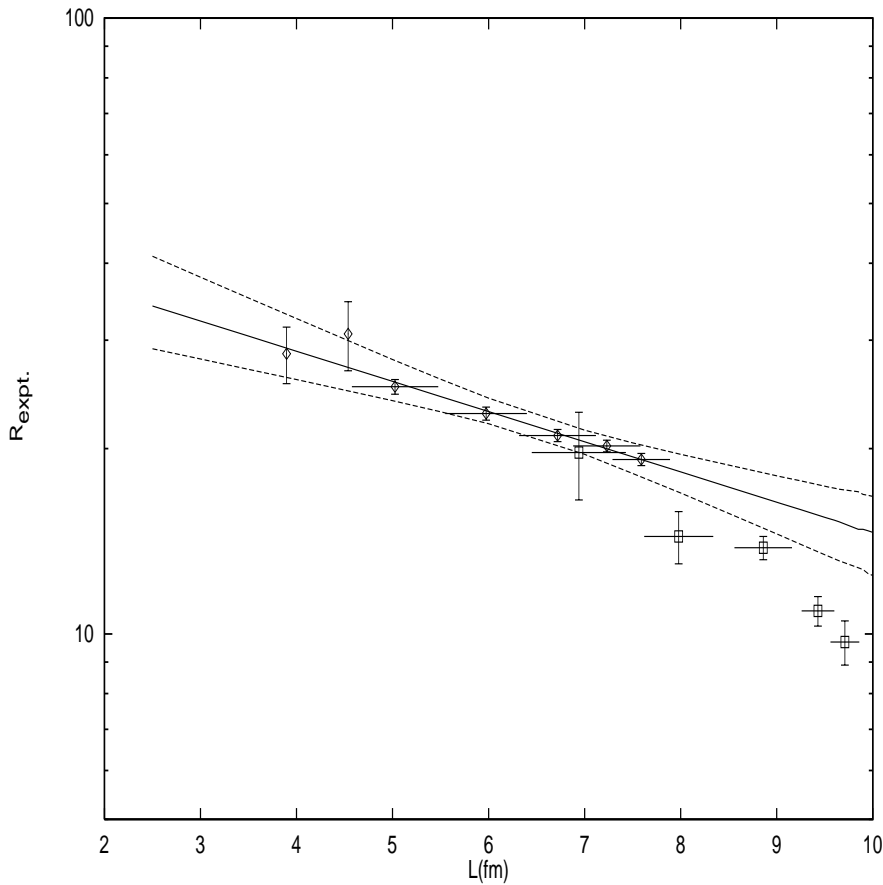}
\caption{ The ratio of $J/\psi$ cross section and the Drell-Yan
cross section vs. $L$ in fm.  The diamonds are NA38 data, shown along
with the straight line fit, a 2$\sigma$ band around it, and the
NA50 data (squares) with 2$\sigma$ errors on them. } 
\end{center}\end{figure}

\begin{figure}[htbp]\begin{center}
\epsfig{height=12cm,width=12cm,angle=-90,file=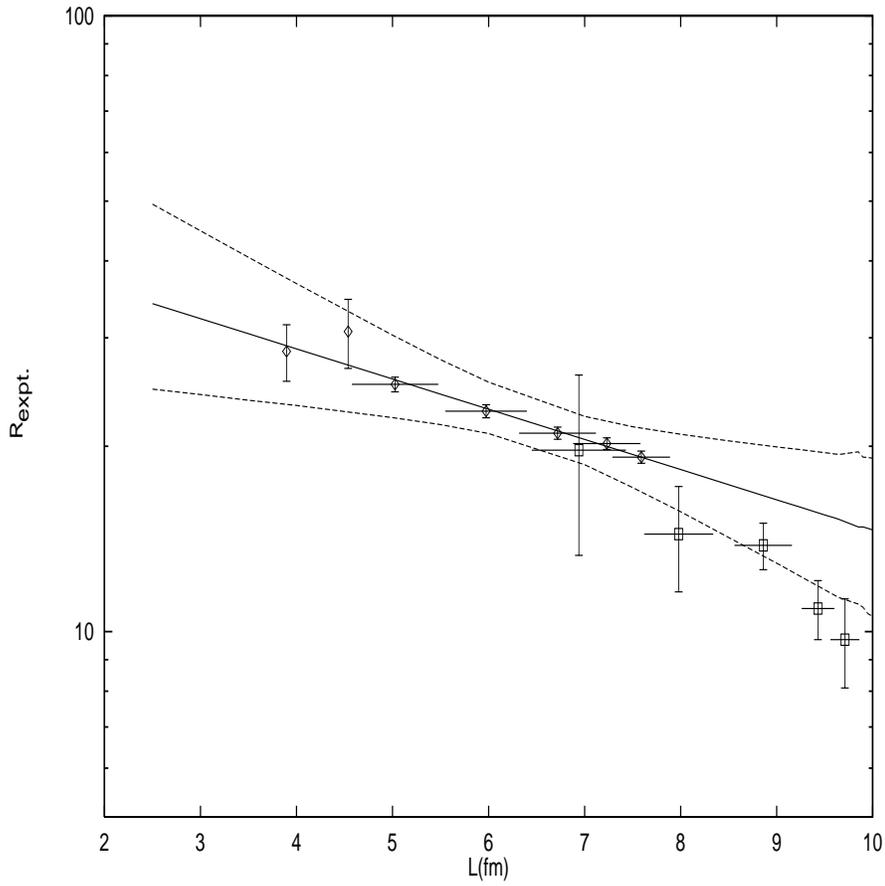}
\caption{ Same as Fig. 1 but with a 4$\sigma$ band around the
fit and the NA50 data with 4$\sigma$ errors on them. } 
\end{center}\end{figure}

Fig. 1 compares the fit (\ref{fitl.vals}) with the NA38 $pA$ and 
$S{\rm -}U$ data.  A 2$\sigma$ band around the fit is shown along 
with the NA50 $Pb{\rm -}Pb$ data with 2$\sigma$ error at each point.  
The band includes the effects 
of the full covariance matrix for the fit parameters in equation
(\ref{fitl.vals}) using the trick mentioned above.  Unlike the
total cross section data, there is a now clear deviation of the
$Pb{\rm -}Pb$ data from the theoretical prediction at a 95 \%
confidence level, especially for the two points at the largest 
$L$-values.   They are compatible with the fit only at
a 4$\sigma$ level, as shown in Fig. 2.  Of course, in both
figures one clearly sees that all the deviations are in the same
direction, which adds further weightage to the observation
of anomalous behavior.  Nevertheless, our results suggest a much
less stronger result, if one goes by the usual wisdom of 
requiring a 5$\sigma$ effect for a new discovery.

It may not be out of place to comment on 1) the origin of our different
result vis-a-vis those in Ref. \cite{Na50Plb} and 2) the implications
for the preliminary 1996 data. In both the cases, discussed above,
Ref. \cite{Na50Plb} compares the ratio of the measured value to 
the theoretical prediction with unity whereas we compare the difference 
of the measured and theoretical cross sections.
Both procedures will give same result provided the theoretical
prediction has no errors at all, which is what seems to be assumed in
Ref. \cite{Na50Plb}. Since the theoretical prediction has big
errors, which stem in one case from large extrapolation and in another
case from the fact that the fitted experimental data has errors in
{\it both} $x$ and $y$ directions, we think our procedure is more
appropriate.  Of course, it is really the neglecting of the propagated 
errors on the predictions which gives rise to a larger result for the 
deviations, i.e., larger apparent anomalous suppression,  than what the 
data warrant.  The preliminary 1996 data\cite{Qm97}
are mostly in agreement with the 1995 data and may be slightly
higher up for the larger $E_T$ or equivalently larger $L$.  Since neither
the $pA$ nor the $S{\rm -}U$ data have changed, none of the fits reported
here change; if the data were available using the same rescaling
as in Ref. \cite{Na50Plb}, one could directly put them on Figs. 1 and 2.
An indirect comparison via the transverse energy plots of Ref. \cite{Qm97} 
yields a preliminary conclusion on the 1996 data that the anomalous 
suppression at any $L$ will unlikely be more than 4$\sigma$ for
them as well.  The same will also hold true for any discontinuity in 
the data, if present.

In conclusion, we have shown that the spread in $L$, which arises
due to the presence of many events in a given in $E_T$ bin with
varying $L$, causes it to be known less precisely than assumed previously.
Taking into account this inevitable imprecision makes the 
theoretical prediction less accurate due to error propagation.  However,
the $L$-dependence of the $J/\psi$ cross section shows a definite
anomalous suppression at the 95 \% confidence level, while 
the $A^\prime A$-dependence of the cross section is not anomalous 
at that level.  On the other hand, the anomalous suppression
at each $L$ is at best a 4$\sigma$ effect.  Better data for
lighter nuclei in finer bins of $E_T$ are needed to improve the 
significance of the $Pb{\rm -}Pb$ results, as the dominant errors 
are in the theoretical predictions.

\bigskip 

I wish to thank R. Burkhalter, H. Shanahan, K. Kanaya and A. Ukawa 
for discussions and persistent questions which led to this work.
The kind hospitality of the all the members of the Center for Computational
Physics, University of Tsukuba, Japan, where this work was done,
is gratefully acknowledged.

\end{document}